\documentclass[10pt]{article}
\usepackage{geometry}
\usepackage{graphicx}
\usepackage{amssymb}
\usepackage{epstopdf}
\usepackage{wrapfig}
\usepackage{amsmath}
\newcommand{\be}{\begin{equation}}
\newcommand{\ee}{\end{equation}}

\long\def\greybox#1{%
    \newbox\contentbox%
    \newbox\bkgdbox%
    \setbox\contentbox\hbox to \hsize{%
        \vtop{
            \kern\columnsep
            \hbox to \hsize{%
                \kern\columnsep%
                \advance\hsize by -2\columnsep%
                \setlength{\textwidth}{\hsize}%
                \vbox{
                    \parskip=\baselineskip
                    \parindent=0bp
                    #1
                }%
                \kern\columnsep%
            }%
            \kern\columnsep%
        }%
    }%
    \setbox\bkgdbox\vbox{
        \pdfliteral{0.9 0.9 0.9 rg}
        \hrule width  \wd\contentbox %
               height \ht\contentbox %
               depth  \dp\contentbox
        \pdfliteral{0 0 0 rg}
    }%
    \wd\bkgdbox=0bp%
    \vbox{\hbox to \hsize{\box\bkgdbox\box\contentbox}}%
    \vskip\baselineskip%
}

\title{Surprises in numerical expressions of physical constants}
\author{Ariel Amir, Mikhail Lemeshko, Tadashi Tokieda}

\begin{document}
\date{}
\maketitle

\begin{abstract}
In science, as in life, `surprises' can be adequately appreciated only in the presence of a null model,
what we expect {\it a priori}.  In physics, theories sometimes express the values of dimensionless physical constants as combinations of
mathematical constants like $\pi$ or $e$.  The inverse problem also arises, whereby the measured
value of a physical constant admits a `surprisingly' simple approximation in terms of well-known
mathematical constants.   Can we estimate the probability for this
to be a mere coincidence, rather than an inkling of some theory?   We answer the question in
the most naive form.
\end{abstract}

   How do we react to the observation that the ratio of the proton mass to the electron mass equals
$6\pi^5$?  This was the content of \cite{Lenz}, reprinted here in full.

\begin{figure}[h]
   \centering
   \includegraphics[width=3.6in]{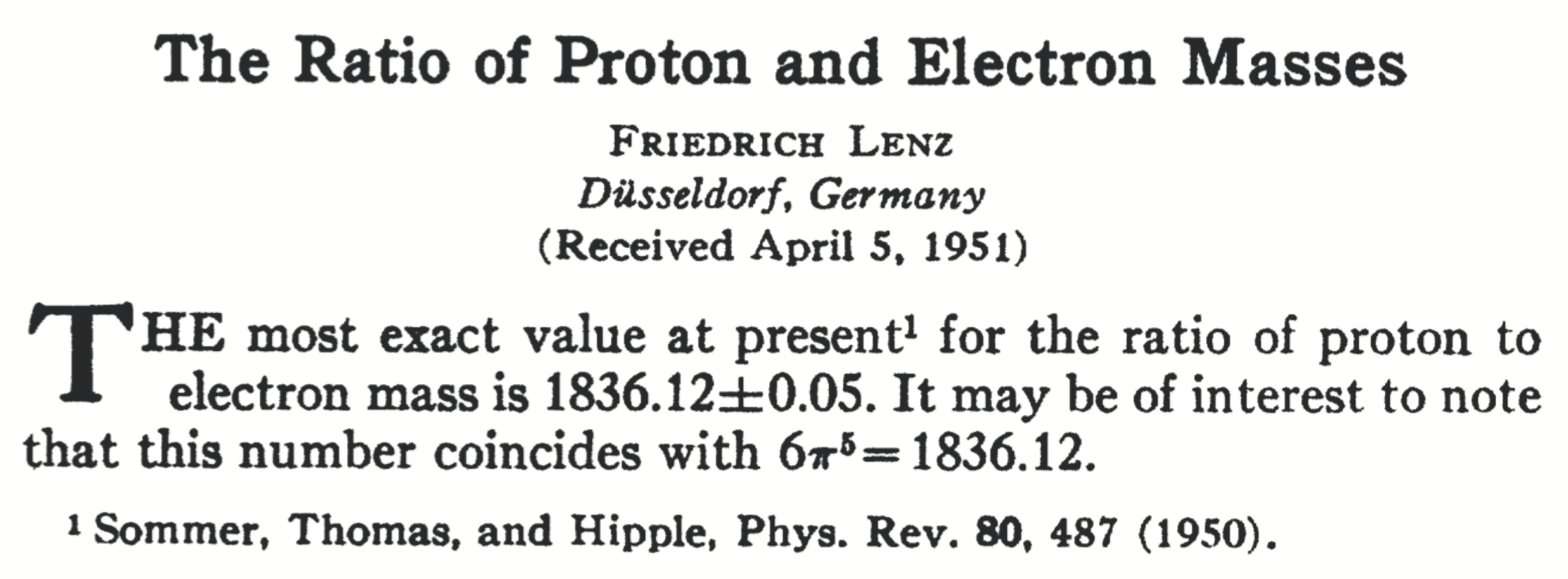}
\end{figure}

\noindent The {\it parvum opus\/}  breathed new life into the ancient field of applied numerology in
fundamental physics (e.g.\ \cite{Wyler1}, \cite{Wyler2}, \cite{Schwartz}; see also an interesting paper \cite{Cover}),
besides adding a 3rd Lenz to the history of physics (after Heinrich, of induced currents, and Wilhelm, of
the conserved vector).  As of 2015 the most exact value of the proton-to-electron mass ratio is
$1836.15267245(75)$ according to the National Institute of Standards and Technology
\cite{NIST}, which still coincides with $6\pi^5 = 1836.1181\cdots$ to 5 significant figures.
How surprising is it that a dimensionless physical constant coincides to such an accuracy with a simple
expression built up from well-known mathematical constants?

    Occasionally, a coincidence is no mere fluke, but rather signals a consequence of some (perhaps as yet undiscovered) theory.  At absolute temperature $T$, a black body
has total emissive power $\sigma T^4$, with
$$
\sigma = 40.802624638\cdots \times \frac{k^4}{h^3 c^2}
$$
(the Stefan-Boltzmann law).  The combination of Boltzmann's constant $k$, Planck's constant $h$, the speed of light $c$
is imposed by dimensional analysis.  But who ordered the funny prefactor?  It coincides
with $2\pi^5/15$ to 7 figures \cite{NIST}.  On this occasion, there turns out to be a theory
due to Planck, and the apparent coincidence is a theorem:
$$
40.802624638\cdots = \frac{2\pi^5}{15} =  2\pi \Gamma(4) \zeta(4) = 2\pi \int_0^{\infty} \! \frac{x^3}{e^x - 1} \, {\rm d}x.
$$

   The question is, how can we distinguish when a significant-looking expression for a physical constant is a fluke, versus when it signals something?  Surely the first step is to figure out the probability that it is a fluke.

      Let us adopt a naive model. We call a {\it $d$-digit number} a number of the form
$$
z_1 z_2\cdots z_{d-1}\, .\, z_d
$$
i.e.\ with $d-1$ digits in its integer part followed by $1$ digit after the decimal point.  Thus, $1836.1$
is a $5$-digit number.  We regard an expression $Z$ as a {\it simple expression built up from well-known
mathematical constants\/} when it is of the form
\begin{equation}
Z = g( f(A, B), C). \tag{*}
\end{equation}
As mathematical constants we allow
$$
A, B, C \in \bigl\{ \> 1, 2, 3, 4, 5, 6, 7, 8, 9, \pi, e, \gamma, \phi, -\widehat{\phi} \> \bigr\}
$$
where $\pi$ and $e$ are uncontroversially well-known,
$\gamma = \lim_{n \to \infty} \bigl( \sum_{k = 1}^n \frac{1}{k} - \log_e n \bigr) = 0.5772\cdots$
is the Euler-Mascheroni constant, $\phi = \frac{1}{2}(1 + \sqrt{5}) = 1.6180\cdots$ is the golden ratio,
$-\widehat{\phi} = - \frac{1}{2}(1 - \sqrt{5}) = 1/\phi = 0.6180\cdots$ is the reciprocal of the
golden ratio.  (Note in passing that $\phi$ is nearly the number of kilometers in 1 mile, which is 1.609344.)
As building operations we allow
$$
f(x, y), g(x, y) \in \bigl\{ \> x + y, \> x - y, \> x \times y, \> x/y, \> x^y, \> \sqrt[y]{x}, \> \log_x y \> \bigr\}.
$$
We can build $6\pi^5$ as $g( f(A, B), C )$ by taking $f(x, y) = x^y, g(x, y) = x\times y$, and $A = \pi, B = 5, C = 6$.

    We generated all expressions $Z$ of the form (*) and, for each $d$, $1 \leqslant d \leqslant 6$, counted the
proportion (probability) they occupy in the set of all $d$-digit numbers.

\begin{figure}[h]
   \centering
   \includegraphics[width=3.7in]{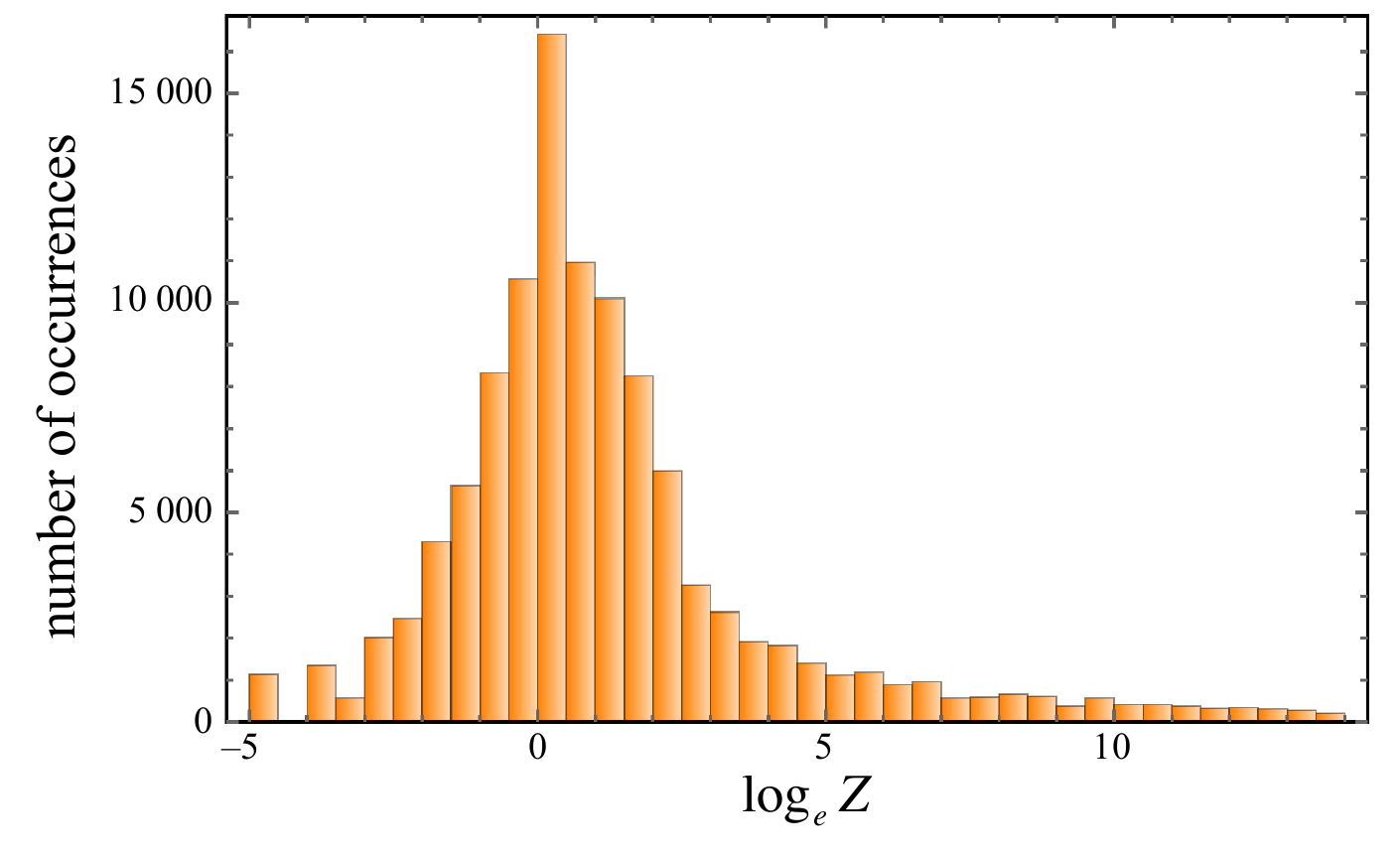}
   \caption {Histogram of log of all possible expressions generated by (*).}
   \label{dist}
\end{figure}

   The histogram of Fig.~\ref{dist} shows the distribution of $\log_e Z$.  It bears a resemblance to a
normal distribution, which suggests that $Z$ itself has a log-normal distribution.
Log-normal distributions govern the product of many independent positive random variables.  Our
$A, B, C$ are not quite random, independent, or many, nor are our $f, g$ necessarily products.  Nonetheless
with optimism we may deem it plausible that a log-normalesque behavior emerges from (*).  Some
of the deviations from log-normal are imputable to our having allowed subtraction, division, exponentiation,
etc.: the sharp peak at $0$ was caused by these operations which, compared with multiplication,
tend to overproduce $e^0 = 1$; they may also cause $Z$ to be negative, in which case we just shrugged and plotted
${\rm Re} \log_e Z$.  Furthermore, we tested the distribution of the leading digits in our data set: it
followed closely Benford's law, where $z$ $(1 \leqslant z \leqslant 9)$ appears with
frequency proportional to $\log(z + 1) - \log z$, e.g.\  1 appeared as the leading digit in about 35\% of
the values of $Z$ and 9 in about 5\%.  Benford's law holds for the leading digits if the log data
distribute uniformly across many orders of magnitude, which is consistent with the log-normal scenario.

\begin{figure}[h]
   \centering
   \includegraphics[width=2.5in]{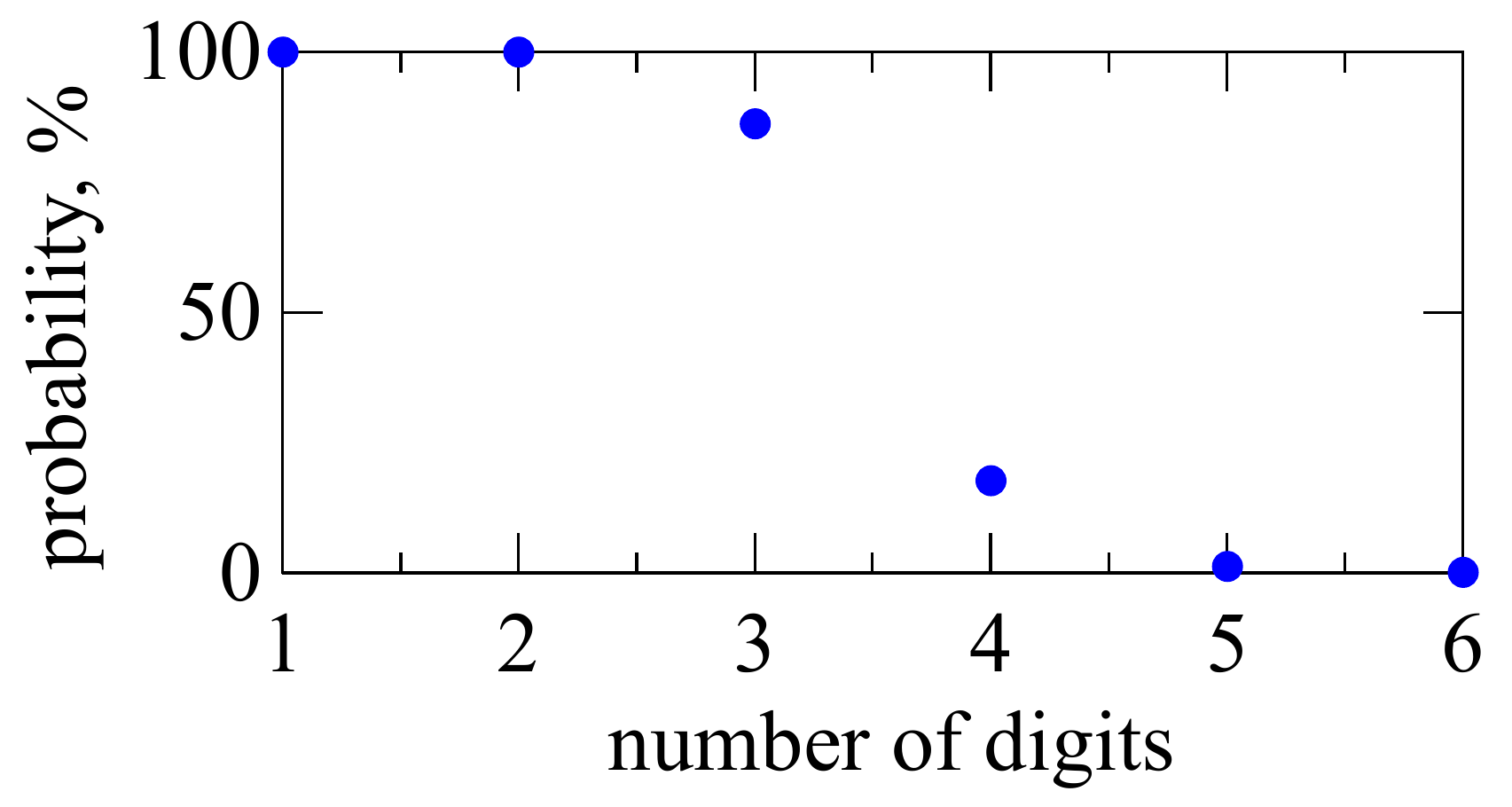}
   \caption{Probability that a random $d$-digit number is expressible by (*), as a function of $d$.}
   \label{plot}
\end{figure}

   The graph of Fig.~\ref{plot} shows the probability that a randomly chosen $d$-digit number coincides
to all its digits with one of the values of $Z$, as we increase $d = 1, \ldots, 6$.
The probability begins at 100\% for $d = 1, 2$, hovers around 86\% for $d = 3$, and tails off
for $d > 4$.  The tail makes heuristic sense.  The number of possible expressions allowed
by (*) is $14^3 \times 7^2 = 10^{5.1\cdots}$.  Therefore $d = 5$ would give an upper
bound for the 100\% coverage if all these expressions were distinct and fitted within the correct magnitude range.
But as witnessed by Fig.~\ref{dist}, $Z \approx 10^{5 - 1}$ or equivalently $\log_e Z \approx 9.2\cdots$ is
already in the tail of the distribution.  In other words, the majority of values of $Z$ overlap and squeeze
into the range of $4$ digits, leaving 5- and 6-digit numbers uncovered.  More diligently, we tally
about 2000 occurrences between $\log_e 10^{4-1} = 6.9\cdots$ and $\log_e 10^{5-1}$,
so that the probability of getting a 5-digit number should be $\frac{2000}{10000 - 1000} \times \frac{1}{10}
\approx 2.2\%$, with the $\frac{1}{10}$ factor because we must take care of the digit after the decimal point.
An actual count yields about 1.2\%.

   We find that the {\it a priori\/} probability for a $5$-digit number like $1836.1$ is only 1.2\%.
So  \cite{Lenz} should surprise us, at least mildly.  Conversely, a simple mathematical expression
that gives a 3- or 4-digit dimensionless physical constant need not get us excited.

   Expanding the reserve of mathematical constants dramatically raises the coverage; so does iterating
operations more times.  For instance, when we bring in constants named after a bevy of
mathematicians as in \cite{wiki}, the coverage for $5$-digit numbers rises to 42\%.  Nowadays some physicists use
an online tool \cite{consts} that mines an immense reserve of constants, in order to solve the inverse problem of
matching the simplest mathematical expression against a given numerical value.  However, it demands a very
high-precision knowledge of the constants, and
when we input $1836.1$, the tool overlooked $6 \pi^5$ and returned much more complicated `solutions'
such as $Y_0(\frac{151}{16}) \times 10^4$.  (This is the unique opportunity to publish the pun: `Why is the Bessel
function of the second kind of order zero denoted by $Y_0$?'  `Why nought?')

    There are other examples of numerical coincidences, from the calendar.  (i) How many seconds are there in 42 days?
The answer is $10!$ exactly.  (ii) How many seconds are there in 1 year ($\approx 365 + \frac{1}{4}$ days)?
The answer is close to $\pi \times 10^7$, with error $< \frac{1}{2}\%$ ;  $\sqrt{10} \times 10^7$ is an even
better approximation, especially in a leap year when the coincidence persists to 5 figures.  Of course,
these examples are disappointing because the numbers depend on the human unit of `second' (whereas
`day' and `year' are natural units).   See \cite{comic} for additional examples.

   A dual issue is addressed by the familiar puzzle of the type `Can we make $n$ (some number thrown off
as a challenge) by using just three 2s?'  This particular puzzle has a universal solution, often attributed to
von Neumann \cite{Halmos}: every integer $n$ can be written
$$
n = - \log_2 \log_2 \!\!\!\!\!\! \underbrace{\sqrt{ \cdots \sqrt{2} } }_{\text{$n$-fold square roots}} \!\!\!\!\! .
$$
The solution milks the notational accident that $\sqrt{x}$ stands for $x^{1/2}$, though if we denote $\log_2$
by $\lg$ as computer scientists do, then we can erase 2 from the base of logarithm and so compensate for
the appearance of 2 in the iterated powers, again getting away with just three 2s.

\medskip

   All three authors contributed equally little to the present study.

\bigskip

\small{
\noindent {\tt arielamir@seas.harvard.edu}

\noindent School of Engineering and Applied Sciences, Harvard University, 29 Oxford St, Cambridge MA 02138, USA
\smallskip

\noindent {\tt mikhail.lemeshko@ist.ac.at}

\noindent IST Austria, Am Campus 1, 3400 Klosterneuburg, Austria
\smallskip

\noindent {\tt tokieda@stanford.edu}

\noindent Department of Mathematics, Stanford University, Stanford CA 94305-2125, USA
\smallskip
}


\begin{thebibliography}{99}
\bibitem{Cover} T.~M.~Cover, On determining the irrationality of the mean of a random variable, {\it Ann.\ Stat.} {\bf 1}
(1973) 862--871.
%\bibitem{Flowers} J.~L.~Flowers \&\ B.~W.~Petley, Progress in our knowledge of the fundamental constants of physics,
%{\it Rep.\ Prog.\ Phys.} {\bf 64} (2001) 1191--1246.
\bibitem{Halmos} P.~R.~Halmos, {\it Problems for mathematicians, young and old}, Mathematical Association of America, 1991.
\bibitem{Lenz} F.~Lenz, The ratio of proton and electron masses, {\it Phys.\ Rev.} {\bf 82} (1951) 554.
\bibitem{Schwartz} H.~M.~Schwartz, On Wyler's derivation of the fine-structure constant, {\it Lett.\ Nuovo Cimento\/}, {{\bf 2} (1971) 1259--1260.}
\bibitem{Wyler1} A.~Wyler, L'espace sym\'etrique du groupe des \'equations de Maxwell, {\it C.\ R.\ Acad.\ Sci.} s\'erie~A
{\bf 269} (1969) 743--745.
\bibitem{Wyler2} A.~Wyler, Les groupes des potentiels de Coulomb et de Yuwaka, {\it C.\ R.\ Acad.\ Sci.} s\'erie~A {\bf 272}
(1971) 186--188.
\bibitem{NIST} {\tt http://physics.nist.gov/cuu/Constants/index.html}
\bibitem{comic}{\tt http://www.xkcd.com/1047}
\bibitem{wiki} {\tt https://en.wikipedia.org/wiki/Mathematical\texttt{\char`_}constant}
\bibitem{consts} {\tt https://isc.carma.newcastle.edu.au}

\end{thebibliography}
\end{document}